\documentclass{ws-ijmpd}

\begin{document}

\markboth{Wagner}
{Physics Insights from Recent MAGIC AGN Observations}

%
\catchline{}{}{}{}{}
%

\title{PHYSICS INSIGHTS FROM RECENT MAGIC AGN OBSERVATIONS}

\author{ROBERT WAGNER on behalf of the MAGIC COLLABORATION}

\address{Max-Planck-Institut f\"ur Physik, F\"ohringer Ring 6, D-80805 Munich, Germany\\
robert.wagner@mpp.mpg.de}

\maketitle

\begin{history}
\received{Day Month Year}
\revised{Day Month Year}
\comby{Managing Editor}
\end{history}

\begin{abstract}
The total set of the 14 active galactic nuclei detected by MAGIC so far
includes well-studied bright blazars like Mkn 501, the giant radio galaxy M 87,
but also the distant flat-spectrum radio quasar 3C 279, and an intriguing
gamma-ray source in the 3C 66A/B region, whose energy spectrum is not
compatible with the expectations from 3C 66A. Besides scheduled observations,
so far MAGIC succeeded in discovering TeV gamma rays from three blazars
following triggers from high optical states. I report selected highlights from
recent MAGIC observations of extragalactic TeV gamma-ray sources, emphasizing
and discussing the new physics insights the MAGIC observations were able to
contribute.
\end{abstract}

\keywords{Active galactic nuclei; BL Lacertae objects; radio galaxies; gamma-rays: observations; gamma-ray telescopes}

\section{Introduction}
Except for the radio galaxies M\,87 and Centaurus A, all 27 currently known
very high energy (VHE, $E\gtrsim 70$ GeV) $\gamma$-ray emitting active galactic
nuclei (AGNs) are blazars,\cite{padovani} assumed to have a close alignment of
their highly relativistic outflows (jets) with our line of sight.  Their
spectral energy distributions (SEDs) are characterized by a synchrotron peak at
optical to X-ray energies, and a second peak at GeV to TeV energies. The latter
can be either due to inverse-Compton (IC) emission from accelerated electrons,
which up-scatter synchrotron photons to high energies.\cite{mar} In hadronic
models, instead, interactions of the jet outflow with ambient matter,
proton-induced cascades, or synchrotron radiation off protons, are responsible
for the high-energy photons.\cite{hadr} Another defining property of blazars is
the high variability of their emission. For some VHE $\gamma$-ray blazars, also
correlations between the X-ray and $\gamma$-ray emission were established on
various time scales.\cite{Wagner4} For sources with a large angle between the
jet and the line of sight (e.g., the radio galaxy M~87), classical IC scenarios
cannot account for the VHE $\gamma$-ray emission. Models that depend less
critically on beaming effects are required instead.\cite{Neronov,Tavecchio}

\section{The MAGIC Telescope}
MAGIC\cite{magic04} is currently the largest (17m-\O) single-dish imaging air
Cerenkov telescope. Its energy range spans from 50--60~GeV up to tens of
TeV. MAGIC is sensitive to $\sim$~1.6\% of the Crab nebula flux in 50 observing
hours. Operation during moderate moonshine extends its duty cycle
substantially,\cite{jrico} which is particularly important for observing
variable $\gamma$-ray sources like AGNs. 
A second MAGIC telescope\cite{magic2} is going to be operational from April 2009 on. By
stereoscopic observations, the sensitivity of MAGIC phase II is increased
considerably.

\section{Strong Flaring of the Radio Galaxy M\,87 in February 2008}
MAGIC detected a strong 9.9-$\sigma$ signal from M\,87  from 2008 January 30 to
2008 February 11.\cite{MAGICM87} The highest flux (15\% that of the Crab
nebula) was recorded on 2008 February 1 at a significance of 8.0 $\sigma$.

Our analysis revealed no variability in the 150--350 GeV range, while a
variable (significance: 5.6\,$\sigma$) $\gamma$-ray flux above 350 GeV on
night-to-night basis confirmed the $E>730\,\mathrm{GeV}$ variability reported
earlier.\cite{hessm87} The variability restricts the VHE emission region to a
size of $R\leq \Delta t\,c\,\delta = 2.6 \times 10^{15}\,\mathrm{cm}$ (Doppler
factor $\delta$), and suggests the core of M\,87 rather than HST-1, the
brightest knot in the M\,87 jet, as the origin of the $\gamma$-rays.  The rapid
variability alone cannot dismiss HST-1 as possible origin of the TeV photons.
In early 2008, however, HST-1 was at a low X-ray flux level,
whereas the luminosity of the M\,87 core was at a historical maximum.\cite{Colin} 
To further elucidate the location of the VHE emission,
the H.E.S.S., VERITAS, and MAGIC collaborations performed an extensive campaign
on M\,87 in 2007/8.\cite{Colin}

For the first time, the energy spectrum below 250 GeV could be assessed. The
hardness of the spectrum, described by a power law with a spectral index
$\alpha = -2.30 \pm 0.11_\mathrm{stat} \pm 0.20_\mathrm{syst}$ is unique among
the VHE AGNs, which show either curved or softer spectra.  No high-energy
cut-off was found, but a marginal spectral hardening, which may be interpreted
as a similarity to the VHE blazars, where such hardening has often been
observed.\cite{Wagner3}

\section{Energy-Dependent Delay in the 2005 July 9 Flare of Mkn 501}
During a high-flux phase of Mkn 501 in summer 2005, this blazar revealed rapid
flux changes with doubling times as short as 3 minutes or less.\cite{alb07a} 
%
For the 2005 July 9 flare, a time delay (zero-delay probability of $P=0.026$)
between the flare peaks at different $\gamma$-ray energies of $\tau_l =
(0.030\pm0.012)\,\mathrm{s\,GeV}^{-1}$ towards higher energies was
found.\cite{alb07b}
Such a delay can be produced by a gradual acceleration of the underlying
electron population.\cite{alb07a} Other explanations are the observation of the
initial acceleration phase of a relativistic blob in the jet\cite{bw08} or,
within an SSC scenario, a brief episode of increased particle
injection.\cite{mas08} An energy-dependent speed of photons in
vacuum,\cite{mat} as predicted in some models of quantum
gravity,\cite{gac,waqg} can also result in a delay: When assuming a
simultaneous emission of the $\gamma$-rays (of different energies) at the
source, a lower limit of $M_{\mathrm{QG}} > 0.21 \times 10^{18}$~GeV (95\%
c.l.), among the best  from time-of-flight measurements,\cite{waqg,fermi} can
be established,\cite{alb07b} which increases further if any delay towards
higher energies in the source is present. 
\section{Blazars Detected during Optical Outbursts}
Observations following high optical states of potential VHE blazars have proven
very successful\cite{OpticalHDGS} with the detection of Mkn 180, 1ES 1011+496,
and recently S5\,0716+71. The 18.7-h observation of 1ES 1011+496 was triggered
by an optical outburst in March 2007, resulting in a 6.2-$\sigma$
detection\cite{MAGIC1011} at $F_{>200\mathrm{GeV}} =
(1.58\pm0.32)\times10^{-11} \mathrm{cm}^{-2} \mathrm{s}^{-1}$. An indication
for a long-term optical--VHE correlation is given, in that in spring 2007 the
VHE $\gamma$-ray flux was $>40$\% higher than in spring 2006, when MAGIC had
observed this blazar for the first time.\cite{systHBL}

In April 2008, the supporting optical KVA telescope reported a bright optical
state of the blazar S5~0716+71, triggering VHE observations, which resulted in
a 6.8-$\sigma$ detection, corresponding to a flux of $F_{>400\mathrm{GeV}}
\approx 10^{-11} \mathrm{cm}^{-2} \mathrm{s}^{-1}$. The source was also in a
high X-ray state.\cite{atel1495giommi}

\section{MAGIC J0223+430: A Gamma-Ray Source in the 3C 66A/B Region}
MAGIC observed the region around 3C~66A for 54.2\,h in fall 2007, resulting in
the discovery\cite{3c66ab} of the $\gamma$-ray source MAGIC~J0223+430 at (RA,
dec) = ($2^{\mathrm{h}} 23^{\mathrm{m}} 12^{\mathrm{s}}$, $43^{\circ} 0.\!'7$),
coinciding with the catalog position of 3C~66B, a nearby ($z=0.022$)
Fanaroff-Riley type I radiogalaxy.  3C~66A, $6.\!'1$ away from the detected
excess, can be excluded statistically as being the $\gamma$-ray source with a
probability of 95.6\%. When allowing for a systematic pointing uncertainty of
$2'$, the exclusion probability is 85.4\%.  The $\gamma$-ray flux was found to
be constant on the level of 2.2\% the Crab nebula flux. The energy spectrum of
MAGIC~J0223+430 extends from 75 GeV to 3 TeV, following a power law with a
photon index of $\alpha=-3.10 \pm 0.31_{\mathrm{stat}} \pm
0.2_{\mathrm{syst}}$. Given the likely association of MAGIC~J0223+430 with
3C~66B, our detection would establish radio galaxies as a new VHE $\gamma$-ray
source class.
In view of the recent detection of 3C~66A as VHE $\gamma$-rays
emitter,\cite{3c66a} we note that if 3C~66A was emitting $\gamma$-rays in 2007
August to December, then its flux was at a significantly lower level than in
2008.  Also it cannot be excluded\cite{TavGhis} that the observed spectrum is a
combination of emission from 3C~66B (dominating at $E>150$\,GeV) and 3C~66A (at
lower energies).  Given the measured spectral index, in the improbable case
that the {\it total} signal originates from 3C~66A, the redshift of this blazar
is likely to be significantly lower than previously assumed\cite{3c66ab}
 ($z=0.444$ and 0.321).

\section{Detection of the Distant Flat-Spectrum Radio Quasar 3C\,279}
VHE $\gamma$-rays were detected from 3C~279 at a 5.77-$\sigma$ post-trial
significance on 2006 February 23, supported by a marginal signal on the
preceding night.\cite{MAGICScience} A zero-flux lightcurve of the full 10-day
observation can be rejected on the 5.04-$\sigma$ level. This makes 3C 279 at
$z=0.536$ the most distant VHE  $\gamma$-ray source.  
The observed VHE spectrum can be described by a power law with a differential
spectral index of
$\alpha=4.1\pm0.7_\mathrm{stat}\pm0.2_\mathrm{syst}$ between $75-500$\,GeV.
It is sensitive to the extragalactic background light\cite{hauser}
(EBL) between $0.2-2\,\mu\mathrm{m}$. The reconstructed intrinsic spectrum is
difficult to reconcile with models predicting high EBL densities (e.g., the
fast-evolution model in Ref. \refcite{stecker}), while low-level
models\cite{primack,ahaebl} are still viable. Assuming a maximum intrinsic
photon index of $\alpha^\ast = 1.5$, an upper EBL limit was
inferred,\cite{MAGICScience} leaving a small allowed region for the EBL.  The
results support, at higher redshift, indications\cite{ahaebl} that {\it Hubble
Space Telescope} and {\it Spitzer} data correctly estimate most of the
background light in the Universe.  


\end{document}